\documentclass[prd, twocolumn, nofootinbib, floatfix]{revtex4-1}

\usepackage{amsmath}
\usepackage{graphicx}
\usepackage{dcolumn}
\usepackage{bm}
\usepackage{epsfig}
\usepackage{amssymb,latexsym,mathrsfs}
\usepackage{graphicx}
\usepackage{color}
\usepackage{hyperref}

\hypersetup{
    colorlinks=true,
    linkcolor=red,
    citecolor=blue,
} 

\newcommand{\be}{\begin{equation}}
\newcommand{\ee}{\end{equation}}
\newcommand{\bs}{\begin{split}} 
\newcommand{\bea}{\begin{eqnarray}}
\newcommand{\eea}{\end{eqnarray}}

\newcommand{\om}{\Omega_m}

\begin{document}

\title{On the Use of Fast Radio Burst Dispersion Measures as 
Distance Measures} 

\author{Pawan Kumar${}^{1}$, Eric V.\ Linder${}^{2,3}$, } 
\affiliation{${}^1$Department of Astronomy, University of Texas, Austin, TX 78712, USA\\
${}^2$Berkeley Center for Cosmological Physics \& Berkeley Lab, 
University of California, Berkeley, CA 94720, USA\\ 
${}^3$Energetic Cosmos Laboratory, Nazarbayev University, 
Nur-Sultan, Kazakhstan 010000}

\begin{abstract} 
Fast radio bursts appear to be cosmological signals whose frequency-time 
structure provides a dispersion measure. The dispersion measure is a 
convolution of the cosmic distance element and the electron density, and 
contains the possibility of using these events as new cosmological distance 
measures. We explore the challenges of extracting the distance in a 
robust manner, and give quantitative estimates for the systematics control 
needed for fast radio 
bursts to become a competitive distance probe. 
The methodology can also be applied to assessing 
their use for mapping electron density 
fluctuations or helium reionization. 
\end{abstract}

\date{\today} 

\maketitle

\section{Introduction}

Fast radio bursts (FRB) are microsecond emissions of radio signals, 
detected with power in the tens of milli-Jansky or greater range 
\cite{lorimer07,thronton13,spitler14,petroff16,shannon18,chime19,ravi19}. 
Several 
dozen FRB are known, with eleven giving repeated signals consisting of 
more than one hundred pulses altogether \cite{spitler16,askap19,chime19,chime19b}. The repeat nature seems to disfavor 
cataclysmic origins such as compact object collisions, and instead FRB 
are generally regarded as radiation emitted from a compact object with a 
high magnetic field, such as a magnetar 
\cite{katz16,katz18,kumar17,lu18,lyutikov17,metzger17,1810.05836}. 

FRB show a characteristic sweep in frequency with time, interpreted as a 
dispersion of the signal propagating through plasma, and the degree of sweep 
and hence dispersion is quantified by the dispersion measure (DM). At 
these radio frequencies, the DM depends on the path length through the 
ionized medium and the electron density. Typical DM of FRB can be several 
hundred to a few thousand in units of parsecs/cm$^3$. This is greater 
than expected from lines of sight within the Milky Way galaxy, plus 
three FRB have been localized sufficiently well to identify their 
host galaxies and measure their 
distances \cite{spitler16,tendulkar17,bannister19,ravi19b}, so FRB are 
regarded as originating at cosmological distances. 

FRB are expected to be useful probes of astrophysics. In particular, they
could provide a good map of the baryon distribution, magnetic fields and 
turbulence in the IGM, helium reionization, and might serve as a new tool for 
measuring cosmological distances to high redshifts $z\gtrsim1$, e.g.\  \cite{macquart13,zhang14,mcquinn14,zhou14,macquart15,walters18}. 
A new cosmological distance measure would be exciting, and useful, if it 
could be made sufficiently robust and competitive in precision to existing 
cosmological distance probes such as Type Ia supernovae standard candles 
and baryon acoustic oscillations standard rulers. Many of the systematic 
uncertainties would differ from other distance probes and thus provide an 
important crosscheck. 

We explore the possibility of FRB serving as a robust distance probe. 
We emphasize that the purpose of the article is not to claim that they 
can do so, but rather to bring together the FRB and cosmology communities 
so they appreciate quantitative estimates of the systematics control that 
would be required, in a fairly general form. 
In Sec.~\ref{sec:method} we describe the relation of dispersion measure 
to distance measure, the comparison with other distance probes, and some 
of the systematics challenges. Section~\ref{sec:sysx} investigates the 
influence of the systematics and quantifies how tightly they need to be 
controlled in order to give competitive cosmological constraints. We 
summarize and discuss future avenues of research in Sec.~\ref{sec:concl}.

\section{From Dispersion Measure to Distance Measure} \label{sec:method} 

The dispersion measure does not directly give the distance to the FRB but 
rather an integral of the distance elements weighted by the electron density. 
That is, 
\be 
{\rm DM}=\int_{t_e}^{t_o} c\,dt\,n_e(t)\,(1+z(t))^{-1}\ , \label{eq:dmdt} 
\ee 
where $t_{e/o}$ is the time of emission/observations, $dt$ is the time 
interval or proper distance, $n_e$ is the (proper) electron number density, 
and the final factor of $1/(1+z)$, where $z$ is the redshift, arises from 
transformations of the frequency and time between the emitter and observer 
frames. Specifically, 
the signal at a given frequency $\nu$ is 
delayed in plasma, with respect to travel time in vacuum, by $dt/\nu^2$ 
(all quantities measured 
in the local frame), and hence in the observer frame the contribution to 
the arrival time delay, the DM, scales as $dt/(1+z)$. 
See for example \cite{0309200,0309364}.

\subsection{General properties} \label{sec:genprop} 

There are several aspects of Eq.~(\ref{eq:dmdt}) to note: 
\begin{itemize} 
\item DM is not purely cosmological but receives contributions from the 
electron density in the burst environment and host and observer galaxies. 
\item If the density weighting were perfectly known, DM is a distance 
combination, say $\int dt/a^2$, that is not trivially related to the 
luminosity distance or angular diameter distance, both of which depend 
on $\int dt/a$, or the time/age interval $\int dt$, where $a=1/(1+z)$ is 
the cosmic expansion factor. 
\item DM is not a pure distance measure but a density weighted sum of 
distance intervals, i.e.\ a convolution of functions. 
\end{itemize} 

While the second item represents a special cosmological opportunity, the 
first and third pose challenges to the use for cosmological model constraints. 
We examine them in order.

\subsection{Component Contributions} \label{sec:component} 

It is useful to start with the first bullet and write the DM in more 
generality with dependence on observational characteristics. As stated in 
the bullet and the literature, DM receives several contributions: 
\bea  
DM_{\rm obs}&=&DM_{\rm env}(\hat n_{\rm env},t)+DM_{\rm host}(\hat n_{\rm host})\notag\\ 
&\qquad&+DM_{\rm cos}(\hat n_{\rm cos},\vec\theta)+DM_{\rm MW}(\hat n_{\rm MW})\ . \label{eq:dmall} 
\eea  
The first term is the dispersion measure induced in the local vicinity 
directly by the fast radio burst source, e.g.\ the plasma associated with 
the physical mechanism and its immediate environment. This DM contribution 
may depend on the orientation $\hat n_{\rm env}$ of the signal propagation 
with respect to, say, the source rotation axis. There may also be a time 
dependence among repeat pulses as the line of sight intersects different 
magnetic field or plasma regions. Both it and the host galaxy term contain 
the same $1/(1+z)$ factor from the observer frame transformation as 
Eq.~(\ref{eq:dmdt}), though we do not explicitly list a redshift dependence 
above. 

The next term comes from the host galaxy and the fourth term comes from 
our own Milky Way galaxy, each possibly depending on where the FRB is 
in the sky and within its host galaxy. The third term is what we seek, 
the cosmological contribution that depends on the cosmic distance of the 
source, which in turn depends on a set of cosmological model parameters 
$\vec\theta$. In addition, it can have a direction dependence due to the 
imperfectly homogeneous intergalactic medium. 

Let us discuss these individually. DM$_{\rm MW}$ can vary from about 
ten to several hundred along a typical line of sight, 
but we will assume that maps of electron 
density and DM in our Milky Way galaxy are sufficiently good that given 
a FRB sky position we can subtract the contribution DM$_{\rm MW}$. 
See, for example, \cite{1610.09448,ne2001}. 
The host galaxy contribution is more problematic. Given an estimate of 
the host galaxy morphology and mass, one could assign a value to 
DM$_{\rm host}$, but $\hat n_{\rm host}$ may not be clear, nor may the 
degree of variation with line of sight for that galaxy type. For the 
Milky Way, DM can vary between lines of sight by over two orders of 
magnitude 
\cite{1610.09448,ne2001}, with the Galactic center magnetar, SGR J1745-2900, 
having a measured of DM=1778 \cite{eatough13}. 
One might hope that this scatter could average to an unbiased mean over 
many FRB host galaxies, but this is not guaranteed, especially if there 
are observational selection effects preferring certain lines of sight 
(e.g.\ perpendicular to a disk plane or see \cite{macquart15b}). 
Presumably this is less of an issue for radio signals than in the optical, 
unless FRBs are 
preferentially located in the galactic center regions or embedded in 
molecular clouds or supernovae remnants. We will keep open the possibility 
of a potential bias in the mean. 

For the local environment contribution DM$_{\rm env}$ we have similar issues 
with directionality. Again, we will allow the possibility that a mean 
over many FRB at the same redshift may not give an unbiased DM estimate 
to that redshift. We do not include time variation as written in 
Eq.~(\ref{eq:dmall}); see Appendix~\ref{sec:local} for further discussion. 

Finally, let us address spatial variations directly in the cosmological 
contribution to DM. Due to inhomogeneities in the intergalactic medium 
(IGM) and halos along the line of sight, 
i.e.\ fluctuations in electron density, different lines of sight to the 
same redshift will have different DM. Simulations 
\cite{1309.4451,1812.11936} indicate the standard deviation in 
DM$_{\rm cos}$ may be 200, i.e.\ 20\%, at $z=1$ (also see 
\cite{1712.01280,prochaska19}). A reasonable approximation 
out to $z\approx3$ is $\sigma_{\rm DM}\approx 210\sqrt{z}$. If one 
approximates DM$\approx 1000\,z$ then 
\be 
\frac{\sigma_{\rm DM,cos}}{\rm DM_{cos}}\approx \frac{20\%}{\sqrt{z}}\ . 
\ee

\subsection{Cosmological Sensitivities} \label{sec:sens} 

The second bullet of Sec.~\ref{sec:genprop} notes the unique 
nature of FRB DM with respect to the distance element dependence. If 
all other issues could be controlled, DM could potentially offer 
complementarity with standard distance probes. Let us explore this aspect 
in further detail. 

If the DM$_{\rm cos}$ contribution could be reliably estimated, and if 
the electron density follows the homogeneous universe dependence of 
$n_e(z)\sim(1+z)^3$, then the integral in DM$_{\rm cos}$ has the form 
$\int dt/a^2=\int dz\,(1+z)/H(z)$, where $H(z)$ is the Hubble parameter. 
This differs from the standard luminosity 
or angular diameter distance form of $\int dt/a=\int dz/H(z)$ and so offers 
different sensitivities to the cosmological parameters for observations 
at various redshifts. In particular, one might expect greater sensitivity 
at higher redshift than in the standard distance case, as well as possible 
complementarity with standard distance probes. 

\begin{figure}[tbp!]
\includegraphics[width=\columnwidth]{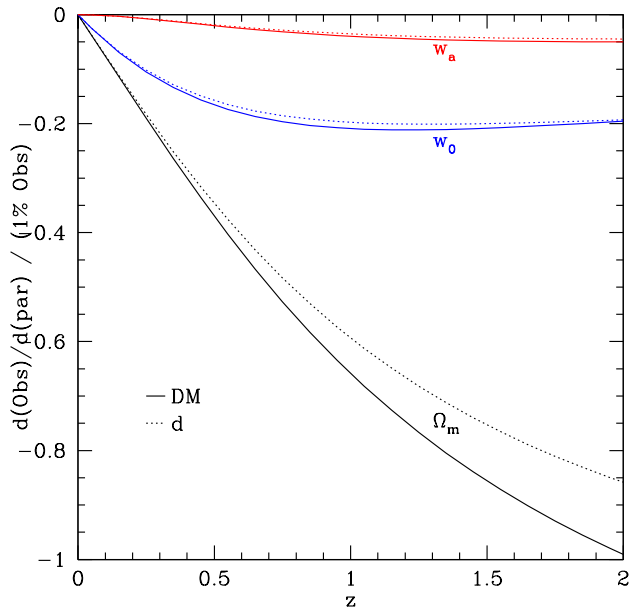}
\caption{
The sensitivity of FRB DM and standard probe distance $d$ to the 
cosmological model parameters are plotted for 1\% measurements of 
the observables at redshift $z$. Very little difference in sensitivity 
is seen, indicating not much complementarity between the different types 
of observations is expected at the cosmological level (although the 
different systematics entering the probes could make the combination 
useful). 
} 
\label{fig:sens} 
\end{figure}

We compute the Fisher sensitivities 
$\partial{\rm Observable}/\partial{\rm parameter}$ for the observable 
being either the FRB DM or the standard distances measured by, e.g., 
supernovae or baryon acoustic oscillations, and the cosmological parameters 
being $\om$, the matter density as a fraction of the critical density, 
and the dark energy equation of state parameters $w_0$ (its value today) 
and $w_a$ (a measure of its time variation). The result are shown in 
Fig.~\ref{fig:sens}.

The extra $1+z$ factor in the integrand for DM makes little difference 
(of course the sensitivity denominator, here at 1\% of the observable 
goes up as well), 
in part because the dark energy parameters have the most impact at low 
redshifts, during cosmic acceleration. For the matter density, there is 
more of an effect, where indeed the sensitivity is somewhat increased 
(at the 10\% level at $z=1$, 15\% level at $z=2$). Since the shapes of 
the curves remain similar, we do not expect significant complementarity. 

\begin{figure}[tbp!]
\includegraphics[width=\columnwidth]{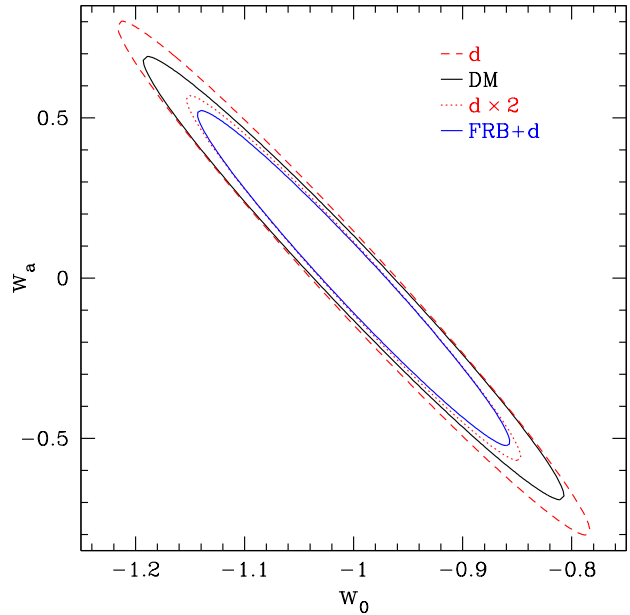}
\caption{
68\% joint confidence contours on the dark energy parameters $w_0$ and 
$w_a$ from measurements of standard distances $d$, or FRB DM, or of both 
(FRB+$d$) to the same redshifts. We also plot the constraints from twice 
as many standard distance measurements ($d\times2$), to give the same 
number of data points as the FRB+$d$ case. All instances include a Planck 
CMB distance prior. 
} 
\label{fig:fisfrb} 
\end{figure}

To quantify this, we calculate the Fisher information matrix for 15 
measurements of 1\% precision from $z=0$--1.5 of DM, of standard 
distance $d$ (it does not matter if it is luminosity or angular diameter 
distance), and of the combination. In particular, the last, in addition to two 
separate data sets, could involve 
simultaneous measurements of the FRB DM and a coincident gravitational 
wave signal (and hence standard siren distance), if the physical mechanism 
giving rise to the FRB also produced detectable gravitational wave signals. 
Also see \cite{1401.0059,margalit19} for FRB and gamma ray bursts. 
Figure~\ref{fig:fisfrb} illustrates the comparison between the different 
cases.

We see that the DM form, if it can be cleanly deconvolved to attain 
DM$_{\rm cos}$ and if it can achieve the same accuracy as standard 
distance measurements, indeed has a slight advantage over conventional 
distance probes. It determines $\om$, $w_0$, $w_a$ to 9\%, 11\%, 14\% 
better, and the dark energy figure of merit (area of the confidence 
region) is 24\% improved. We also notice that the orientation of the 
confidence contour is slightly rotated with respect to the standard 
distance case, indicated some complementarity. If we add together the 
information from FRB DM and standard distances, there is some slight 
improvement over simply using standard distances with the same number 
of measurements, or equivalently total measurement precision. The 
figure of merit for the FRB+$d$ case is 12\% greater than for the 
$d\times2$ case. 

Of course the big question is whether FRB DM can 
be used as a precision distance probe without incurring systematics 
that bias the results. We examine this in the next section.

\section{Systematics and Required Control} \label{sec:sysx} 

If we misestimate the Milky Way, host galaxy, or FRB environment 
contributions to the DM then we will offset the cosmological contribution 
which we may want to use as a distance measure to derive cosmological 
constraints. Similarly, if we model the electron density function 
incorrectly this also adds a systematic uncertainty to the distance. 

We can use the Fisher information bias formalism (see, e.g., 
\cite{knox98,linbias}) to compute the effect of small systematic offsets 
propagating to biases on cosmological model parameters. In particular, we 
will focus on the matter density and dark energy parameters, $\om$, $w_0$, 
$w_a$, usually constrained by distance probes. 
The bias on parameter $p_i$ is given by 
\be 
\delta p_i=\left(F^{-1}\right)_{ij}\sum_k \frac{\partial{\mathcal O_k}}{\partial p_j}\frac{1}{\sigma_k^2}\Delta{\mathcal O_k}\ , 
\ee 
where $\Delta{\mathcal O_k}$ is the difference between the true observable 
and the one with systematics, $F$ is the Fisher information matrix (and so 
$F^{-1}$ is the covariance matrix), $\sigma_k$ is the statistical uncertainty 
on the observation, and the index $k$ sums over observations, e.g.\ redshift 
shells. 

Once the bias $\delta p_i$ is determined, one can then impose some 
requirement on its magnitude as a function of the statistical error on 
that parameter $\sigma(p)$. For example, limiting $\delta p<0.5\sigma(p)$ 
is one approach. This, however, does not take into account correlations 
between parameters, such that a modest bias in multiple parameters in 
the wrong direction can shift the cosmology well outside the true joint 
confidence contour, e.g.\ perpendicular to the major axis of the contours 
in Fig.~\ref{fig:fisfrb}. Therefore we also consider the change in 
likelihood due to the bias, using \cite{shapiro,shapiro2} 
\be 
\Delta\chi^2={\bm{\delta p\,F}}^{(r)}\,{\bm{\delta p}}^T\ , 
\ee 
where $F^{(r)}$ is the Fisher information matrix in the reduced parameter 
space of interest (marginalized over others) and ${\bm{\delta p}}$ is the 
vector of parameter biases. Here the space of interest is three dimensional, 
over $\om$, $w_0$, $w_a$. 

For the observational offsets in the DM cosmology contribution we write 
\be 
{\rm DM_{cos}}=b_a(z)+b_m\int \frac{dz\,(1+z)}{H(z)/H_0}\,(1+z)^n\ . 
\ee 
Here $b_a$ is an additive shift from misestimating other components of DM, 
e.g.\ DM$_{\rm env}$. It can be redshift dependent because of observational 
uncertainties that might depend on, e.g., fluence signal to noise, or 
detector response with frequency as the source redshift gives rise to 
different observer frame frequencies. We model this simply as 
$b_a(z)=B(1+z)^m$. The fiducial value is $B=0$, i.e.\ no misestimation. 

The quantity $b_m$ is a constant multiplicative shift, e.g.\ from 
misestimated amplitude of baryon density $\Omega_b h^2$, metallicity, 
or ionization state, 
with all redshift dependence absorbed into the $(1+z)^n$ factor. Note that 
$b_m$ also has the inverse Hubble constant $1/H_0$ absorbed. Its fiducial 
value is $b_m=1$. We take a simple smooth factor $(1+z)^n$ factor to 
represent all redshift dependence other than that 
already accounted for in $n_e(z)\sim (1+z)^3$, i.e.\ deviations from that, 
including average patchiness $\delta n_e/n_e$, and metallicity etc.\ 
trends. The fiducial value is $n=0$. When all systematic shifts take 
their fiducial values, then there is no misestimation and no cosmology bias. 

We are now set to study how tightly the systematic shifts must be 
constrained in order not to produce a significant cosmological bias from 
the data. Beginning with the additive shift, we find that a redshift 
independent shift ($m=0$) has no effect on the cosmology parameters since 
it is purely an amplitude offset. (Note that it would affect such 
parameters as the physical baryon density $\Omega_b h^2$ or the Hubble 
constant $H_0$, which we do not focus on.) This is equivalent to the 
situation with a change in the Hubble constant for supernova distances: 
that only affects the supernova $\mathcal{M}$ amplitude parameter, not 
the cosmology parameters. 

Next we allow the amplitude shift to vary with redshift, i.e.\ $m\ne0$. 
Specifically $b_a(z)=B_{\rm mid}[(1+z)/(1+z_{\rm mid})]^m$ so $B_{\rm mid}$ 
is the amplitude at the pivot $z_{\rm mid}$ where this is taken to be the 
midpoint of the redshift range of observations, i.e.\ $z_{\rm mid}=0.75$ 
for FRB covering $z=0$--1.5. 
The cosmology bias on each parameter will scale linearly with the pivot 
amplitude, but have a more complicated dependence on the power law index 
$m$. This is because each cosmology parameter feeds differently into the 
distance interval at various redshifts; this is also evident in the Fisher 
sensitivities plotted in Fig.~\ref{fig:sens}. While $\om$ enters strongly 
already at low redshifts, and $w_0$ reaches near maximum impact near the 
midpoint redshift, the observable is only sensitive to $w_a$ at high 
redshift. Thus we expect little bias in $w_a$ for $m<0$ (which weights 
low redshifts) and more for $m>0$ (weighting high redshifts). 

\begin{figure}[tbp!]
\includegraphics[width=\columnwidth]{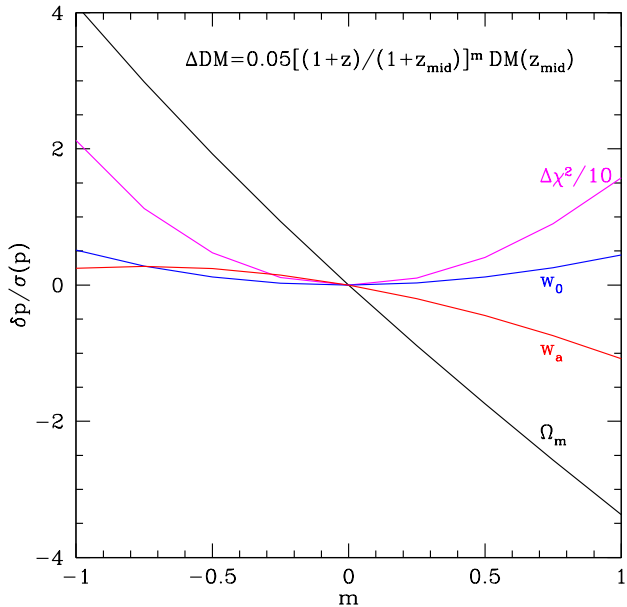}
\caption{
The effect of an additive bias $b_a(z)$, due to misestimation of the 
noncosmological contributions to DM, on the cosmological parameters is 
plotted vs its redshift power law index $m$. A redshift independent additive 
bias only affects the DM amplitude, not the main cosmology parameters. 
Note that $\Delta\chi^2$ is plotted divided by 10. 
} 
\label{fig:biasaddm} 
\end{figure}

Figure~\ref{fig:biasaddm} plots the results for an additive systematic 
of 5\% of the DM value at the midpoint redshift; recall the results will 
scale linearly with amplitude. The biases on the cosmology parameters 
indeed follow the above physical expectations. We see that for a 5\% 
systematic, the bias in $\om$ can reach $4\sigma$, and near $1\sigma$ on 
$w_a$. Furthermore, the best fit cosmology is biased by up to 
$\Delta\chi^2\approx20$ for $|m|$ up to one. For a Gaussian joint confidence 
contour in three parameters, $\Delta\chi^2\approx20$ corresponds to nearly 
$4\sigma$.

Thus, if we want to restrict the cosmology bias to less than 
$0.5\sigma$, say, we need to be able to control the additive systematic 
to $\sim0.6\%$ if $m=1$. Since the bias on $\om$ dominates, and this appears 
from Fig.~\ref{fig:biasaddm} to be fairly linear in $m$, we can phrase the 
requirement as 
\be 
B_{\rm mid}<0.006\,\frac{1}{m}\,\left(\frac{\rm bias}{0.5\sigma}\right)\,
{\rm DM}(z_{\rm mid})\ . \label{eq:addlim} 
\ee 
Since DM$_{\rm cos}(z=0.75)\approx 750$, this requirement implies the 
systematic uncertainties in other contributions to DM (that vary with 
redshift) should be smaller than 
5 pc/cm$^3$ -- a challenging bound. Conversely, if we can control the 
additive systematic to 5\% ($\Delta DM=38$), then we need to ensure that 
the redshift dependence is quite mild, with power law index $m<0.15$. 

For the multiplicative systematic, the scaling of the cosmological 
parameter biases depends both on the amplitude of the DM offset and the 
redshift dependence. That is, 
\be 
\Delta{\rm DM_{cos}}=b_m\int \frac{dz\,(1+z)}{H(z)/H_0}\,\left(\frac{1+z}{1+z_{\rm mid}}\right)^n 
-\int \frac{dz\,(1+z)}{H(z)/H_0}\ . 
\ee 
Only for $n=0$ do the biases scale linearly with 
$(b_m-1)$. Note that when $n=0$, the integral value is still redshift 
dependent and so a constant multiplicative offset does propagate into 
cosmology. 
For $n=0$ the redshift dependence of the electron density is as in the usual 
case, but the amplitude $b_m$ can differ due to, e.g., misestimation of 
the total ionization fraction. 

\begin{figure}[tbp!]
\includegraphics[width=\columnwidth]{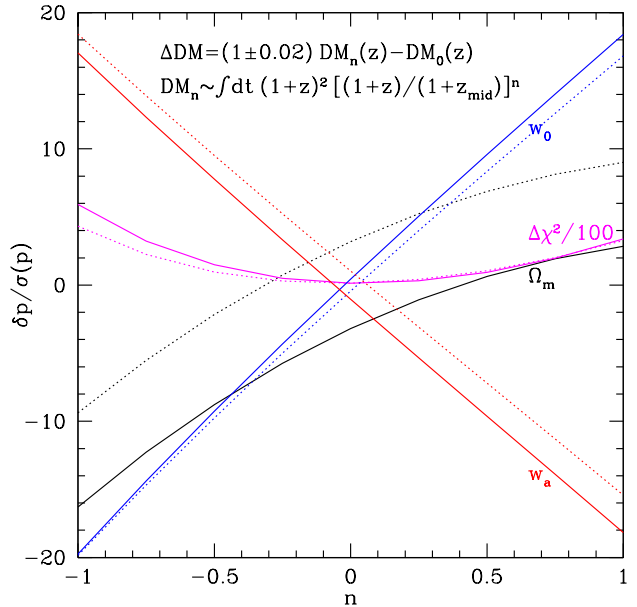}
\caption{
The effect of a multiplicative bias, due to misestimation of the, e.g., 
electron density contributions to DM, on the cosmological parameters is 
plotted vs its redshift power law index $n$. Solid curves are for $+2\%$ 
systematic and dotted curves are for $-2\%$; the biases are not 
antisymmetric about zero bias except for $n=0$. 
Note that $\Delta\chi^2$ is here plotted divided by 100. 
} 
\label{fig:biasmult} 
\end{figure}

Figure~\ref{fig:biasmult} plots the results for a multiplicative 
systematic of $\pm2\%$ (i.e.\ $b_m=1\pm0.02$) of the DM value at the 
midpoint redshift. Since the biases do not scale linearly with this offset 
amplitude (except at $n=0$), the curves are not antisymmetric about zero 
bias. Increasing the systematic further roughly shifts the curves up and 
down, while preserving their shape.

Since for the multiplicative case the systematic DM offset is calculated 
from the difference between the integral with the added $(1+z)^n$ 
redshift scaling and the standard case with $n=0$, $b_m=1$ then even a 
parameter like $w_a$ more sensitive at high redshift shows comparable 
effects whether $n=+1$ or $n=-1$. In fact, the biases involve a complicated 
combination of the covariances from all the parameters in the inverse 
Fisher matrix $F^{-1}$. 

For the $n=0$ case, the $\Delta\chi^2$ offset in the joint cosmological 
parameter confidence contours is about 14, which corresponds to $3\sigma$ 
in three dimensions. This increases for larger $|n|$, reaching 
$\Delta\chi^2=600$ for $n=-1$ for a 2\% misestimation. Thus, severe 
requirements must be placed on systematic uncertainties of both $b_m$ and 
$n$ in order to ensure robust cosmological conclusions. 

At $n=0$, the analog of Eq.~(\ref{eq:addlim}) for the multiplicative 
systematic is 
\be 
|b_m-1|<0.003\,\,\left(\frac{\rm bias}{0.5\sigma}\right)\ . 
\label{eq:multlim0} 
\ee 
For $b_m=1$, the constraint on the redshift power law index is 
\be 
|n|<0.027\,\,\left(\frac{\rm bias}{0.5\sigma}\right)\ . 
\label{eq:multlimbm1} 
\ee 
In this regime $w_0$ and $w_a$ are the parameters most sensitive to bias. 
A value of $n=0.027$ corresponds to a 1\% deviation from the standard 
homogeneous behavior of $(1+z)^3$ in the electron density at $z=1.5$ and 
a $-1.5\%$ deviation at $z=0$. 

Note that the requirements on systematics control would tighten further 
if we combine systematic 
uncertainties, e.g.\ allowing both $b_m\ne1$ and $n\ne0$, or allowing both 
additive and multiplicative systematics.

\section{Discussion and Conclusions} \label{sec:concl} 

Fast radio bursts are an intriguing astrophysical phenomena that provides 
a new backlight on the intergalactic medium, and potentially a new 
distance measure. We investigated the relation between the observed 
dispersion measure and the cosmological distances, pointing out three 
main elements: 1) we must understand the noncosmological contributions 
to DM, and their possible variations with direction, redshift, and time, 
2) FRB provide a unique distance measure that is not trivially related 
to standard luminosity or angular distances, and 3) the convolution of 
electron density and cosmic distance needs to be considered (see 
Appendix~\ref{sec:deconvolve} for some possibilities). 

The main focus was on the first point. 
We quantified the effects of systematic uncertainties in extraction of 
$DM_{\rm cos}$ from the observed DM on the cosmological parameters, 
and found that they could be quite substantial.  
The amplitude of additive systematics (noncosmological contributions 
to DM) should be under 0.6\% and multiplicative systematics (e.g.~electron 
density variation) smaller than 0.3\% in order that the bias in the cosmological
parameters is less than $0.5\sigma$ (scaling formulas were provided). 
Moreover, the redshift dependence 
of these effects, i.e.\ power law indices, should be estimated 
to better than $\delta m<0.15$ and $\delta n<0.027$. 
These are going be very challenging measurements, even averaging over 
many FRB, since there could be selection effects. 

For example, the DM for the Milky Way varies over $\sim20$--2000 for different
lines of sights through the Galaxy.
It is entirely possible that the host contribution to DM could vary similarly,
and that might introduce a bias with host galaxy orientation. 
If there is a systematic bias that is not reduced by $\sqrt{N}$, 
this would make it very hard to subtract the host galaxy contributions to DM
and obtain a robust estimate of DM$_{\rm cos}$. 

Clearly, for FRB dispersion measures to be used as distance measures, and 
useful probes of cosmology, systematics will need careful attention and 
control. There are some hopeful signs. 

Upcoming FRB surveys such as \cite{shannon18,chime19} will provide
good localizations so that the redshifts for a fraction of these 
bursts would be determined from host galaxy identification. These FRBs 
could better provide the DM contribution from the local environment of bursts 
and their host galaxies, hopefully reducing the systematics 
in DM$_{\rm cos}$. 

Another source of information on distances is contained in pulse 
broadening. Microsecond scale fluctuations in FRB lightcurves (LCs) 
are smoothed out due to radio waves scattering off of electron density 
fluctuations in the IGM and the interstellar medium (ISM) 
of the host galaxy and the Milky Way. The broadening of pulses in the FRB LC
due to scatterings increases with distance to the source,
and this effect is more pronounced at lower frequencies; the scattering width 
scales as $\sim \nu^{-4.4}$. Thus, measurements of the smallest time
variation of the FRB LCs can provide information to help improve 
the determination of DM$_{\rm cos}$. The contributions to the 
FRB pulse broadening by the FRB host galaxy ISM 
and the Milky Way ISM are
suppressed by the geometrical factor $4f(1-f)$ in comparison to the IGM 
scattering, where $f$ is the ratio of the distances between the FRB source
and the host galaxy ISM (treated as a turbulent screen) and the source
and us. Therefore, the contribution of the IGM to the FRB pulse width
broadening can be important in spite of its much lower electron density
compared with the ISM of the host galaxy.

If the IGM turbulence is similar to the Galactic ISM (other than, of course,
the large difference in the electron densities),
then the pulse broadening in time due to IGM density fluctuations can be obtained by 
rescaling the ISM contribution by three orders of magnitude to account for 
the lower electron density in the IGM \cite{lorimer13,caleb16}, 
\be 
\label{wIGM}
w_{\rm IGM} = (4\times 10^{-11}\,{\rm ms})\, 
(1+2\times 10^{-3} DM_{\rm IGM}^{2}) \frac{DM_{\rm IGM}^{2.2}}{\nu_{0,{\rm GHz}}^{4.4}} \ . 
\ee 

An alternate possibility is to use the theoretical
expression for temporal smearing by IGM turbulence, e.g.~\cite{macquart13},
instead of rescaling the Galactic ISM scattering observations, to estimate
FRB pulse broadening, 
\be 
w_{\rm IGM}(z) = \frac{k_{\rm IGM} D_{\rm eff}}{\nu_{0,GHz}^{4}(1+z_L)} 
\int^{z}_{0}\frac{dz^{\prime}\,(1+z^\prime)^3}{H(z^{\prime})/H_0} \ , 
\label{wIGM2}
\ee 
where $D_{\rm eff}=D_L D_{LS}/D_S$, with $D_L$, $D_S$, $D_{LS}$ the 
angular diameter 
distance to the lens, to the source, and between the lens and source, and 
$k_{\rm IGM}=1.18\times10^{13}\ {\rm ms\ MHz^{4}}$ is a normalization factor 
\cite{bhattacharya19}. 

Eventually, when we have direct redshift measurements for a large sample
of FRBs and their pulse widths due to IGM scatterings, then an empirical
relation may be obtained for the IGM turbulence properties and scattering
broadening. The empirical IGM turbulence property together with the DM 
measurement would provide a better distance estimate that could be subject 
to less systematic uncertainty. 

Finally, if the goal is astrophysics, then for a given cosmology, FRBs are 
likely to be excellent probes of electron density fluctuations, and thus the 
baryon density spectrum on large
scales. They may also map the epoch of helium reionization \cite{1902.06981}, or constrain 
the CMB optical depth degeneracy \cite{1901.02418}; all these 
are exciting science topics.

\acknowledgments 

EL is supported in part by the Energetic Cosmos Laboratory and by the 
U.S.\ Department of Energy, Office of Science, Office of High Energy Physics, 
under Award DE-SC-0007867 and contract no.\ DE-AC02-05CH11231.

\appendix 

\section{Local Environment} \label{sec:local} 

Let us briefly consider further the issue of directionality, or variation, 
of the local environment contribution DM$_{\rm env}$. 
One possibility is that then 
a mean over many FRB at the same redshift may not give 
an unbiased DM estimate. There is an element of environmental 
directionality that could be more problematic. A given direction local 
to the FRB engine could potentially have a DM varying over time, due to 
rotation, magnetic field motion, plasma velocity, etc. This is especially 
relevant for repeating  
FRBs; recall that accurate redshifts are most likely to be obtained 
for repeat systems. 

\begin{figure}[tbp!]
\includegraphics[width=\columnwidth]{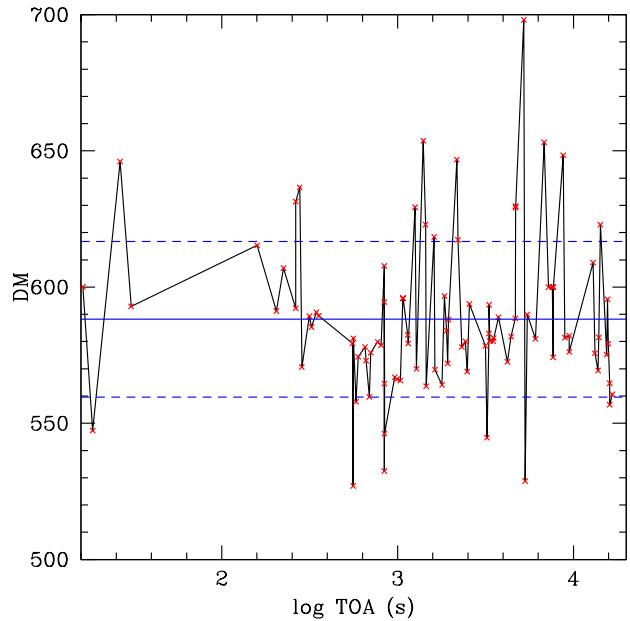}
\caption{
The 93 pulses of the repeating FRB 121102 detected at various times of 
arrival on August 26, 2017 have variation in total DM (red x's). While the 
mean DM=588.2 pc/cm$^3$ (horizontal solid blue line), there is not 
only a standard deviation 28.6 pc/cm$^3$ (horizontal dashed blue lines) 
but significant outliers. This may make it unclear what value of DM$_{\rm env}$ 
to subtract off to obtain the cosmological DM, but see the text regarding the 
DM plotted. 
} 
\label{fig:dmtime} 
\end{figure}

Consider the 93 pulses for a repeat outburst of FRB 121102 detected over 
five hours as described in \cite{1809.03043}. Figure~\ref{fig:dmtime} 
plots the total DM of FRB 121102 vs arrival time, using the data from 
Table~2 of \cite{1809.03043}. 
An important distinction is that these are DM$_{\rm S/N}$, which may not be 
the physical DM, and indeed \cite{1804.04101,1811.10748} advocate use of a 
DM$_{\rm struc}$, which relies on an assumption that substructures 
within a pulse are all emitted at the same time. This may indeed be correct, 
and in the main text we assume it is and that there is no time variation, 
but it is worthwhile considering this a little further here. 
We note that rotation measures can show substantial time 
variation \cite{1801.03965}, and \cite{1804.04101} state they cannot rule 
out large variations in DM between pulses. Furthermore, \cite{1804.04101} 
gives reasons why DM$_{\rm S/N}>$DM$_{\rm struc}$, and yet we see in 
Fig.~\ref{fig:dmtime} significant deviations in DM below as well as above 
the mean.

The following argument suggests this may not be a physical variation. 
If the DM were to vary on a time scale of, say, 5 hours, then that suggests 
that the region responsible for this variation is of size no larger than 
$10^{14}\,$cm (if the medium is moving at much less than the speed of light 
then the size is correspondingly smaller than $10^{14}\,$cm). The density of 
such a medium, in order for it to contribute $\delta{\rm DM}\sim 100$, 
should be $n_e>10^6\,{\rm cm}^{-3}$. However, such a medium would block the 
FRB radiation because of induced-Compton scatterings 
\cite{lu18,blandford75,wilson82,lyubarsky16}. 
Also see \cite{1707.02923}. 
Hence the possibility
of a large DM variation on a timescale of a few hours (or days) seems 
difficult to obtain (though there may be a regime of plasma lensing 
where this can be realized, cf.\ \cite{cordes17}). 

Therefore, while we remain cautious about the 
possibility of not yet fully identified systematics -- merely noting that 
if one used the maximum deviation from the mean seen for one pulse in 
Fig.~\ref{fig:dmtime} to derive DM$_{\rm cos}$ this would give a 50\% 
distance error, and if one took the standard deviation value (dashed lines 
in Fig.~\ref{fig:dmtime}) this would give a 14\% distance error -- 
we will not consider further physical time variations 
in DM$_{\rm env}$.

\section{Deconvolution} \label{sec:deconvolve} 

As an interesting aside, 
we now turn to the third bullet of Sec.~\ref{sec:genprop}. 
Suppose that we can successfully separate out the cosmological contribution
DM$_{\rm cos}$. 
This involves a convolution of the distance element with the electron 
density along the signal propagation path. If one assumes a functional 
form for $n_e(z)$, e.g.\ $n_{e,0}(1+z)^3$ as in a perfectly homogeneous, 
and homogeneously ionized, universe then this is not an issue. Conversely, 
for an arbitrary function of redshift it is not possible to separate out 
the cosmological distance; one sees that for 
\be 
n_{e,2}(z)=n_{e,1}(z)\,\frac{dt_1(z)}{dt_2(z)}\ , 
\ee 
one could mimic, and misinterpret, cosmology model 1 distances as those of 
cosmology model 2, for all redshifts. 

Observables involving products of the distance with other quantities is 
not unusual. For example, for strong lensing time delays $\Delta t$ the 
cosmological distance $D_{\Delta t}$ is entwined with the Fermat factor 
$\Delta\phi$ of the lensing gravitational potential, 
\be 
\Delta t=D_{\Delta t}\,\Delta\phi\ . 
\ee 
This is merely a multiplicative relation, though, not a convolution 
in the integrand, and even so is a major systematic for strong lensing, 
with detailed modeling of the lens matter distribution required for 
accurate extraction of the cosmic distance. 

Closer to the present case is the relation of the observable gamma ray 
flux of annihilating dark matter particles in a galaxy to the dark matter 
properties such as mass and cross section. We can write this illustratively 
as 
\bea 
F_\gamma&\sim& \int dr_{\rm los}\int d^3v_1\int d^3v_2 P(|\vec v_2-\vec v_1|) 
f(\vec r,\vec v_1) f(\vec r,\vec v_2)\\ 
&\sim& P_0\,J[\rho(r)]\ . 
\eea 
Here $f$ is the phase space distribution function of the dark matter 
particles, $P$ represents particle physics properties, and the $J$ factor 
involves an integral over the dark matter density (squared). 
(One can write a similar expression for direct detection of dark matter 
by nuclear recoil, where the analogous astrophysical factor depends linearly 
on the dark matter density.) For an 
arbitrary $J$ factor, i.e.\ dark matter density profile $\rho(r)$, we 
cannot determine all the dark matter properties uniquely. Instead one must 
assume particular models, i.e.\ functional forms, for the density profile. 
For FRB, this corresponds to a model for $n_e(z)$. Any error in the model 
could lead to a systematic error in the cosmic distances, and vice versa. 

In fact, in the dark matter case \cite{1707.07019} there are clever 
mathematical methods for treating a fairly general convolution function. 
These are based on convex hulls and give a piecewise continuous solution, 
but require the function to be monotonic. (Another application of convex 
hulls to eigenvector bounds for an arbitrary dark energy equation of state 
is discussed in \cite{0908.2637}.) While in a homogeneous universe 
the function $n_e(z)$ is indeed monotonic, in this case we already know 
the functional form and do not need to use the theorem. We do not 
particularly expect $n_e(z)$ to be monotonic, as we know the realistic 
intergalactic medium is patchy, as discussed in Sec.~\ref{sec:component} 
and \cite{1309.4451,1812.11936}. Furthermore, $n_e$ contributions to 
DM$_{\rm cos}$ also arise (roughly equal to the IGM contribution 
\cite{1309.4451,1812.11936}) from the line of sight intersecting galaxy 
halos, and this is emphatically nonmonotonic. 

Thus the convex hull 
approach does not seem apt, unless we seek bounds on the cosmic 
contribution. Note that a model independent principal component approach 
to cosmic reionization, e.g.\ \cite{0705.1132}, could be adapted to the 
late time ionization fraction and combined with 
the convex hull eigenvector bound approach of \cite{0908.2637} -- if we 
were interested in bounds on cosmic distances rather than the distances 
themselves.



\begin{thebibliography}{99}

\bibitem{lorimer07}
D.R. Lorimer, M. Bailes, M.A. McLaughlin, D.J. Narkevic, F. Crawford, 
A Bright Millisecond Radio Burst of Extragalactic Origin, Science 318, 777 (2007)

\bibitem{thronton13}
D. Thornton et al., A Population of Fast Radio Bursts at Cosmological Distances, Science 341, 53 (2013)

\bibitem{spitler14}
L.G. Spitler et al., Fast Radio Burst Discovered in the Arecibo Pulsar ALFA Survey,  ApJ 790, 101 (2014)

\bibitem{petroff16}
E. Petroff, E.D. Barr, A. Jameson, E.F. Keane, M. Bailes, M. Kramer, V. Morello,D. Tabbara, W. van Straten, FRBCAT: The Fast Radio Burst Catalogue, PASA 33, e045 (2016)

\bibitem{shannon18}
R.M. Shannon et al., The dispersion-brightness relation for fast radio
bursts from a wide-field survey, Nature 562, 386 (2018)

\bibitem{chime19}
M. Amiri et al., Observations of fast radio bursts at frequencies down to 
  400 megahertz, Nature 566, 230 (2019)

\bibitem{ravi19}
V. Ravi, The observed properties of fast radio bursts, Monthly Notices of Royal
Astronomical Society 482, 1966 (2019)

\bibitem{spitler16}
L.G. Spitler et al., A repeating fast radio burst, Nature 531, 202 (2016) 

\bibitem{askap19}
P. Kumar et al., Faint repetitions from a bright Fast Radio Burst source, arXiv:1908.10026 (2019)

\bibitem{chime19b}
B.C. Andersen et al., CHIME/FRB Detection of Eight New Repeating Fast Radio Burst Sources, arXiv:1908.03507 (2019)

\bibitem{katz16}
J.I. Katz, Fast radio bursts — A brief review: Some questions, fewer answers, Modern Physics Letters A, 31, 1630013 (2016)

\bibitem{katz18}
J.I. Katz, Fast Radio Bursts, Progress in Particle and Nuclear Physics 103, 1, (2018)

\bibitem{kumar17}
P. Kumar, W. Lu \& M. Bhattacharya, Fast radio burst source properties and curvature radiation model, MNRAS 468, 2726 (2017)

\bibitem{lu18}
W. Lu and P. Kumar, On the radiation mechanism of repeating fast radio bursts,
 MNRAS 477, 2470 (2018)

\bibitem{lyutikov17}
M. Lyutikov, Fast Radio Bursts’ Emission Mechanism: Implication from Localization, ApJL 838, L13 (2017)

\bibitem{metzger17}
B.D. Metzger, E. Berger, B. Margalit, Millisecond Magnetar Birth Connects FRB 121102 to Superluminous Supernovae and Long-duration Gamma-Ray Bursts, ApJ 841, 14 (2017)

\bibitem{1810.05836} 
E. Platts, A. Weltman, A. Walters, S.P. Tendulka, J.E.B. Gordin, S. Kandhai, 
A Living Theory Catalogue for Fast Radio Bursts, arXiv:1810.05836 

\bibitem{tendulkar17}
S.P. Tendulkar et al., The Host Galaxy and Redshift of the Repeating Fast Radio Burst FRB 121102, ApJ 834, L7 (2017)

\bibitem{bannister19}
K.W. Bannister, A single fast radio burst localized to a massive galaxy at cosmological distance, Science 365, 565 (2019)

\bibitem{ravi19b}
V. Ravi, A fast radio burst localized to a massive galaxy, Nature 572, 352 (2019)

\bibitem{zhang14}
W. Deng and B. Zhang, Cosmological implications of fast radio burst/GRB association, ApJL 783, L35 (2014)

\bibitem{mcquinn14}
M. McQuinn, Locating the missing baryons with extragalactic dispersion measure estimates, ApJL 780, L33 (2014)

\bibitem{zhou14}
B. Zhou, X. Li, T. Wang, Y-Z Fan, and D-M Wei, Fast radio bursts as a 
cosmic probe?, PhRvD 89, 7303 (2014)

\bibitem{macquart15}
J-P. Macquart et al., Fast Transients at Cosmological Distances with the
SKA, Proceedings of Advancing Astrophysics with the Square Kilometre Array

\bibitem{walters18}
A. Walters, A. Weltman, B. M. Gaensler, Y-Z Ma, and A. Witzemann, 
Future Cosmological Constraints From Fast Radio Bursts, ApJ 856, 65 (2018) 

\bibitem{macquart13}
J-P. Macquart, J.Y. Koay, Temporal Smearing of Transient Radio Sources by the Intergalactic Medium, ApJ 776, 125 (2013) 

\bibitem{0309200} 
K. Ioka, Cosmic Dispersion Measure from Gamma-Ray Burst Afterglows: Probing 
the Reionization History and the Burst Environment, ApJL 598, L79 (2003) 
[arXiv:astro-ph/0309200] 

\bibitem{0309364} 
S. Inoue, Probing the Cosmic Reionization History and Local Environment of 
Gamma-Ray Bursts through Radio Dispersion, MNRAS 348, 999 (2004) 
[arXiv:astro-ph/0309364] 

\bibitem{1610.09448} 
J.M. Yao, R.N. Manchester, N. Wang, A New Electron Density Model for 
Estimation of Pulsar and FRB Distances, ApJ 835, 29 (2017) [arXiv:1610.09448] 

\bibitem{ne2001} 
J.M. Cordes, T.J.W. Lazio, arXiv:astro-ph/0207156 

\bibitem{eatough13}
R.P. Eatough et al., A strong magnetic field around the supermassive black hole at the centre of the Galaxy, Nature 501, 391 (2013) 

\bibitem{macquart15b}
J-P. Macquart, S. Johnston, On the paucity of fast radio bursts at low Galactic latitudes, MNRAS 451, 3278 (2015) [arXiv:1505.05893] 

\bibitem{1812.11936} 
M. Jaroszynski, Fast Radio Bursts and cosmological tests, MNRAS 484, 1637 
(2019) [arXiv:1812.11936] 

\bibitem{1309.4451} 
M. McQuinn, Locating the ``missing'' baryons with extragalactic dispersion 
measure estimates, ApJ 780, L33 (2014) [arXiv:1309.4451] 

\bibitem{1712.01280} 
J.M. Shull, C.W. Danforth, The Dispersion of Fast Radio Bursts from a 
Structured Intergalactic Medium at Redshifts $z < 1.5$, ApJL 852, L11 (2018) 
[arXiv:1712.01280] 

\bibitem{prochaska19}
J.X. Prochaska and Y. Zheng, Probing Galactic haloes with fast radio bursts, MNRAS 485, 648 (2019) [arXiv:1901.11051] 

\bibitem{1401.0059} 
W. Deng, B. Zhang, Cosmological implications of Fast Radio Burst / Gamma-Ray 
Burst Associations, ApJL 783, L35 (2014) [arXiv:1401.0059] 

\bibitem{margalit19}
B. Margalit, E. Berger, and B.D. Metzger, Fast Radio Bursts from Magnetars Born in Binary Neutron Star Mergers and Accretion Induced Collapse, arXiv:1907.00016 (2019)

\bibitem{knox98} 
L. Knox, R. Scoccimarro, S. Dodelson, Impact of Inhomogeneous Reionization 
on Cosmic Microwave Background Anisotropy, Phys. Rev. Lett. 81, 2004 (1998) 
[arXiv:astro-ph/9805012] 

\bibitem{linbias} 
E.V. Linder, Biased cosmology: Pivots, parameters, and figures of merit, 
Astropart. Phys. 26, 102 (2006) [arXiv:0604280] 

\bibitem{shapiro} 
C. Shapiro, Biased Dark Energy Constraints from Neglecting Reduced Shear 
in Weak Lensing Surveys, ApJ 696, 775 (2009) [arXiv:0812.0769] 

\bibitem{shapiro2} 
S. Dodelson, C. Shapiro, M. White, Reduced Shear Power Spectrum, 
Phys. Rev. D 73, 023009 (2006) [arXiv:astro-ph/0508296] 

\bibitem{lorimer13}
D.R. Lorimer, A. Karastergiou, M.A. McLaughlin, \& S. Johnston, On the detectability of extragalactic fast radio transients, MNRAS 436, L5 (2013)

\bibitem{caleb16}
M. Caleb, C. Flynn, M. Bailes, E.D. Barr, R.W. Hunstead, E.F. Keane, V. Ravi, 
 W. van Straten, Are the distributions of fast radio burst properties consistent with a cosmological population?, MNRAS 458, 708 (2016)

\bibitem{bhattacharya19}
M. Bhattacharya, P. Kumar and D. Lorimer, Population modelling of FRBs from intrinsic properties, arXiv:1902.10225 

\bibitem{1902.06981} 
M. Caleb, C. Flynn, B.W. Stappers, Constraining the era of helium reionization using fast radio bursts, 
MNRAS 485, 2281 (2019) [arXiv:1902.06981] 

\bibitem{1901.02418} 
M.S. Madhavacheril, N. Battaglia, K.M. Smith, J.L. Sievers, 
Cosmology with kSZ: breaking the optical depth degeneracy with Fast Radio 
Bursts, arXiv:1901.02418 

\bibitem{1809.03043} 
Y.G. Zhang, V. Gajjar, G. Foster, A. Siemion, J. Cordes, C. Law, Y. Wang, 
Fast Radio Burst 121102 Pulse Detection and Periodicity: A Machine Learning 
Approach, ApJ 866, 149 (2018) [arXiv:1809.03043] 

\bibitem{1804.04101} 
V. Gajjar et al, Highest-frequency detection of FRB 121102 at 4-8 GHz using 
the Breakthrough Listen Digital Backend at the Green Bank Telescope, 
ApJ 863, 2 (2018) [arXiv:1804.04101] 

\bibitem{1811.10748} 
J.W.T. Hessels et al, FRB 121102 Bursts Show Complex Time-Frequency 
Structure, arXiv:1811.10748 

\bibitem{1801.03965} 
D. Michilli et al, An extreme magneto-ionic environment associated with 
the fast radio burst source FRB 121102, Nature 553, 182 (2018) 
[arXiv:1801.03965] 

\bibitem{blandford75}
R.D. Blandford and E.T. Scharlemann, On Induced Compton Scattering by Relativistic Particles, Ap \& SS 36, 303 (1975)

\bibitem{wilson82}
D.B. Wilson, Induced Compton Scattering in Radiative Transfer, MNRAS 200, 
  881 (1982)

\bibitem{lyubarsky16}
Y. Lyubarsky and S. Ostrovska, Induced Scattering Limits on Fast Radio Bursts
   from Stellar Coronae, ApJ 818, 74 (2016)

\bibitem{1707.02923} 
Y-P. Yang, B. Zhang, Dispersion Measure Variation of Repeating Fast Radio 
Burst Sources, ApJ 847, 22 (2017) [arXiv:1707.02923] 

\bibitem{cordes17}
J.M. Cordes,I. Wasserman,J.W.T. Hessels, T.J.W. Lazio, S. Chatterjee,
R.S. Wharton, Lensing of Fast Radio Bursts by Plasma Structures in Host Galaxies, ApJ 842, 35 (2017)

\bibitem{1707.07019} 
G.B. Gelmini, J-H. Huh, S.J. Witte, Unified Halo-Independent Formalism From 
Convex Hulls for Direct Dark Matter Searches, JCAP 1712, 039 (2017) 
[arXiv:1707.07019] 

\bibitem{0908.2637} 
J. Samsing, E.V. Linder, Generating and Analyzing Constrained Dark Energy 
Equations of State and Systematics Functions, Phys. Rev. D 81, 043533 (2010) 
[arXiv:0908.2637] 

\bibitem{0705.1132} 
M.J. Mortonson, W. Hu, Model-independent constraints on reionization from 
large-scale CMB polarization, ApJ 672, 737 (2008) [arXiv:0705.1132] 



\end{thebibliography}
\end{document}